\documentclass[11pt]{article}
\usepackage[linesnumbered,ruled,vlined]{algorithm2e}
\usepackage{array}
\usepackage{amsmath}
\usepackage{amssymb}
\usepackage{graphicx}
\usepackage{amsthm}
\usepackage{latexsym}
\newtheorem{theorem}{Theorem}
\newtheorem{cor}{Corollary}
\newtheorem{definition}{Definition}
\newtheorem{prop}{Proposition}

\newtheorem{example}{Example}

\newcommand{\beq}{\begin{equation}}
\newcommand{\eeq}{\end{equation}}
\newcommand{\barr}{\left[\begin{array}}
\newcommand{\earr}{\end{array}\right]}

\newcommand{\bpf}{\begin{proof}}
\newcommand{\epf}{\end{proof}}

\newcommand{\ftwo}{\ensuremath{\mathbb{F}_{2}}}

\newcommand{\bi}{\begin{itemize}}
\newcommand{\ei}{\end{itemize}}
\newcommand{\bnum}{\begin{enumerate}}
\newcommand{\enum}{\end{enumerate}}
\newcommand{\bc}{\begin{center}}

\begin{document}
\title{Implicant based parallel all solution solver for Boolean satisfiability}
\author{Virendra Sule\\Department of Electrical Engineering\\
Indian Institute of Technology Bombay, Powai\\Mumbai 400076, India\\(vrs@ee.iitb.ac.in)}
\maketitle

\begin{abstract}
This paper develops a parallel computational solver for computing all satifying assignments of a Boolean system of equations defined by Boolean functions of several variables. While there are well known solvers for satisfiability of Boolean formulas in CNF form, these are designed primarily for deciding satisfiability of the formula and do not address the problem of finding all satisfying solutions. Moreover development of parallel solvers for satisfiability problems is still an unfinished problem of Computer Science. The solver proposed in this paper is aimed at representing all solutions of Boolean formulas even without the CNF form with a parallel algorithm. Algorithm proposed is applied to Boolean functions in algebraic normal form (ANF). The algorithm is based on the idea to represent the satisfying assignments in terms of a complete set of implicants of the Boolean functions appearing as factors of a Boolean formula. The algorithm is effective mainly in the case when the factors of the formula are sparse (i.e. have a small fraction of the total number of variables). This allows small computation of a complete set of implicants of individual factors one at a time and reduce the formula at each step. An algorithm is also proposed for finding a complete set of orthogonal implicants of functions in ANF. An advantages of this algorithm is that all solutions can be represented compactly in terms of implicants. Finally due to small and distributed computation at every step as well as computation in terms of independent threads, the solver proposed in this paper is expected to be useful for developing heuristics for a well scalable parallel solver for large size problems of Boolean satisfiability over large number of processors.       
\end{abstract}

\noindent
Category: cs.DS, cs.SC, cs.DC, q-bio.MN\\
ACM class: I.1.2, F.2.2, G.2\\
MSC class: 03G05, 06E30, 94C10

\section{Introduction}
Solving Boolean systems of equations is a core computational problem in fields such as verification of computing hardware, deduction, verification and synthesis of logic,  constraint satisfaction problems, Multivariable Quadratic (MQ) equations over binary field, optimization under logic constraints, cryptanalysis, problems in graph theory etc.\ which arise in Computer engineering, Cryptology, Optimization and Integer programming \cite{crah, hook, scho, bard, ffha}. Since recent times a most glaring new application of solving Boolean systems arises in Biological  (Genetic Regulatory) Networks \cite{honh, alab, guyk, kars}. Two basic versions of such problems are the \emph{decision problem} in which it is required to determine the consistency (or solvability) of the system and, presenting \emph{all solutions} of the problem when the system is consistent. Most general version of such problems involve co-efficients and solutions over general Boolean algebras \cite{rude}. However this general problem has hardly been investigated computationally. On the other hand Boolean systems over the two element Boolean algebra $B_{0}=\{0,1,+,.,'\}$ are most useful for formulation of computational problems of applications in a natural way such as the well known CNF \emph{satisfiability} problem called CNF SAT (solving satisfiable assignment of CNF formulas) \cite{scho} and their applications.

This paper is aimed at solving the problem of the second kind, that of presenting all solutions of Boolean systems over $B_{0}$ when the equations are specified in algebraic normal form (ANF) also called polynomial form with $B_{0}$ co-efficients. For conformity with literature on SAT we shall term this problem as All-SAT problem. Concurrently there are two other problems which can be termed as \emph{minwt-SAT} (respectively \emph{maxwt-SAT}) of producing all solutions with minimum (respectively maximum) Hamming weights of solutions. In fact searching for solutions with minimum or maximum weight of assignments follows solving the problem of presenting all solutions with a specified weight. Hence the problem of presenting all solutions encompasses solutions to all these related problems. In many applications such as Biological networks and Cryptography it is known that solutions exist to the given system but it is important to find all solutions. Since number of solutions can exponentially blow up due to free variable assignments, a compact way to represent all solutions is to represent them in terms of the union of unique assignments. We provide such an approach using the implicants of Boolean functions. 

Problem such as minwt-SAT is motivated by computation of maximal likelihood error patters in error correcting codes. Minimizing a weight associated with a solution of Boolean systems is also a natural problem in optimization under logic constraints. The algorithm proposed in this paper is based on computation of implicants of individual Boolean equations. We show that this approach creates parallel (independent) threads over which the computation can be performed at the cost of spreading computations over parallel threads. This parallel strategy is expected to be scalable for handling special systems of Boolean equations in which individual equations involve sparse Boolean functions so that the load of computing the prime implicants of individual equations is controlled by the sparsity and can be taken as a constant in worst case. Then the computation time in each thread is determined by substitution of partial assignments. However, an important advantage of the use of implicants is that it results in a compact representation of all solutions in terms of partial assignments of variables. This is a vexing problem in satisfiability as the number of solutions blow up exponentially while the implicant representation is the most compact way to represent the blowing up set. This paper is a sequel to earlier efforts at understanding solutions of Boolean equations by the author which are reported in \cite{sul1,sul2,desu,des2}.

\subsection{Elimination theoretic methods}
A representation of all solutions of Boolean equations resulting from elimination of Boolean variables has been well known \cite{rude,brow}. Another way to represent all solutions is the \emph{reproductive form} of solutions which can be used to computationally generate solutions uniquely \cite{rude, crah}. Elimination of variables also leads to represent all solutions compactly through a recursive sequence of inequalities between functions which bound the Boolean values of eliminated variables \cite{rude}. From a computational point of view it would be valuable to understand the scalability of the reproductive form of solutions as well as elimination procedure using parallel computation for handling large systems efficiently. However such studies appear to be absent in the literature. 

Elimination process requires that all equations containing a variable to be eliminated be converted to an equation with a single Boolean function. On the other hand, an inherently parallel computation resorts to as much decomposition and independent processing of equations. The implicant based approach to expressing satisfying assignments leads to such a decomposition and hence is inherently suited to parallel computation. Achieving scalability under parallel computation is a practical challenge of computation which is yet to be satisfactorily resolved in many computational problems. It is for this reason, search for new algorithms for the solution of Boolean equations is of importance.

\subsection{Background of literature}
We shall only consider Boolean systems over $B_{0}$. Hence formal expressions of Boolean functions will have co-efficients and assignments of variables in $B_{0}$. Such a system of $m$ equations can be considered without loss of generality in the form
\beq\label{Boolsys}
f_{i}(x_{1},\ldots,x_{n})=1
\eeq
$i=1,\ldots,N$ where $f_{i}$ are Boolean functions in ANF. The $n$-variables shall be denoted in short by $X$. The system (\ref{Boolsys}) is said to be \emph{satisfiable} in short SAT, if there exists an $x$ in $B_{0}^{n}$ at which each of the Boolean equations when evaluated are satisfied. Otherwise it is called \emph{un-satisfiable} or un-SAT. Thus satisfiability is equivalent to consistency of the single equation
\[
F(X)=\prod_{i=1}^{m}f_{i}(X)=1
\]

In this form we call $F$ the given Boolean formula and $f_{i}$ \emph{factors} of $F$. Among the most common algorithms considered for solving such algebraic systems are the XL method and the Grobner basis algorithm (with variables over the binary field $\ftwo$). These methods have been extensively studied \cite{cidm, bard} and have been generalized to equations over more general finite fields \cite{ffha}. Such approaches, though elegant mathematically, run into practical difficulties in solving large systems as they do not inherently employ decomposition or parallelism. On the other hand parallelization of these algorithms has not proved to be effective for solving systems of industrial sizes as is evident from the vast literature \cite{crah}. Also a major problem in generation of all rational solutions of these equations in which variables take values, the finite field of co-efficients, is that the said algebraic approaches usually result in finding solutions over an algebraically closed field.

Perhaps a simplest approach for solving sparse systems is the search through decomposition as proposed in \cite{yzou} which assumes systems less than a certain small threshold size are easily solvable, decomposes the large system into smaller subsystems which are all solved and the solutions combined to find solutions of the larger system. This approach is thus a heuristic in handling large systems and is the well known message passing protocol in computer science. The patching of solutions of smaller systems in practice is also computationally intensive and involves communication between processing nodes. Hence such methods need careful study of scalability performance. 

The problem of deciding satisfiability of a set of CNFs (CNF-SAT) is also well studied \cite{knut, hand} and has led to quite successful approaches for solving industrial size problems as well as those arising in diverse applications, commonly termed SAT approaches \cite{scho}. However SAT solvers are known to address the decision problem and not the All-SAT problem which requires representation of all solutions. SAT solvers also do not address problems with minimization of cost on solutions, in particular the problem of minwt-SAT. Development of parallel SAT solvers is also largely in infancy. The paper \cite{hamw} considers some of the issues in parallel SAT solvers. Some of the important applications of Boolean systems are comprehensively covered in \cite{crah} along with literature on computational work. It readily follows that a solution to a problem with minimization of cost such as minwt-SAT exists iff the system (\ref{Boolsys}) is SAT though this approach is practically useful only for satisfiable systems as un-satisfiability is decided only at the end of the algorithm.

Purpose of this paper is to address the All-SAT problem of seeking all solutions $x$ of (\ref{Boolsys}) over $B_{0}^{n}$ by representing the solutions in terms of \emph{implicants} of equations $f_{i}=1$. This approach allows inherent parallelization of search and storage and hence has potential to be scalable for addressing industrial size systems.

\section{Complete set of implicants and partial assignments}
In this paper we propose to address the problem of representing all satisfying assignments of Boolean equations in terms of partial assignments specified by implicants of component factors of Boolean formulas. For a Boolean function $f(X)$ in variables $X$ whose assignments are considered over the set $B_{0}^{n}$, our approach shall center around the notion of a complete set of implicants of $f$. There are many such sets each of which is connected with an sum of product (SOP) representation of $f$. Any such SOP contains all the information about satisfying assignments of $f$. Among all such representations the set of prime implicants of $f$ is unique. However the SOP representation of $f$ in prime implicants is not irreducible or minimal, although contains richer information about multiplicity of zeros of equations $f=0$. Prime implicants have played important role in propositional logic \cite{qua1,qua2,brow} while Boolean minimization is important in digital logic \cite{koha}. We begin this section by gathering background on implicants.    

Let $f:B_{0}^{n}\rightarrow B_{0}$ be a Boolean function of $n$-variables ordered as $X=\{x_{1},\ldots,x_{n}\}$ denoted in short as $f(X)$. The set of all such functions denoted $B(n)$ turns into a Boolean algebra under naturally defined sum, product and complement of functions \cite{brow}. Let $A$ denote a subset of indices $\{1,2,\ldots,n\}$ of variables $X$. A term in variables $x_{k}$ for $k$ in $A$ is the product
\[
t=\prod_{k\in A}x_{k}^{\alpha_{k}}
\]
where $\alpha_{k}=0,1$, $x^{0}=x'$ and $x^{1}=x$. $A$ defining the term $t$ is called support of $t$. We may also denote such a term as
\[
t=\prod_{k\in A}l_{k}
\]
where $l_{k}=x_{k}^{\alpha_{k}}$ is called a literal of variable $x_{k}$ when $\alpha_{k}$ is known. 

For any term $t$ as above, the equation $t(X)=1$ has the unique solution assignments of variables denoted as 
\[
(t)=\{x_{k}=\alpha_{k}, k\in A\}
\]
(or $l_{k}=1$ for $k\in A$). However as a Boolean equation $t(X)=1$ considered in all variables $X$, above assignment represents the set of all assignments in $B_{0}^{n}$ in which variables not involved in $t$ are free. Hence, the assignment $(t)$ above is called as \emph{partial assignment} defining the \emph{solution hypersurface} $S(t)$ in $B_{0}^{n}$ of all solutions of the equation $t(X)=1$. Formally
\[
S(t)=\{a\in B_{0}^{n}|t(a)=1\}
\]
is a hypersurface like object defined by $t$ which is the set of all true values of $t(X)$ as a Boolean function. Note that the set $S(t)$ is the same as the \emph{subcube} in $B_{0}^n$ of true values of $t$ which is used for geometric interpretation of true values of Boolean functions represented in DNF \cite[section 1.9]{crah}.

\begin{definition}\emph{
Given a Boolean function $f(X)$ any term $t$ of variables in $X$ such that as functions, $t\leq f$ in the Boolean algebra $B(n)$, is called an \emph{implicant} of $f$. An implicant $t$ is thus equivalently defined as any term $t$ such that $f(S(t))=1$. An implicant $t$ of $f$ is called \emph{prime} if any implicant $q$ such that $t\leq q\leq f$ implies $t=q$. A set of implicants $I(f)$ of $f$ is called \emph{complete} if $f(a)=1$ for an assignment $X=a$, then there exists an implicant $t$ in $I(f)$ such that $t(a)=1$ or alternatively $a\in A(t)$. For two terms $s$, $t$ the term $s$ is said to be \emph{formally included} in $t$ if $s\leq t$ as functions which is equivalent to $s=tr$ for some term $r$. A complete set of implicants $I(f)$ is empty iff $f(X)$ is a zero function while $I(f)=\{1\}$ iff $f(X)=1$ is a tautology.
}
\end{definition}

A notion of \emph{ratio} or \emph{quotient} of Boolean functions has been common in the literature. We shall need ratios of functions only with respect to terms. For a Boolean function $f(X)$ and a term $t$ in $X$, notation $f/t$ for a quotient of $f$ by $t$ shall denote the Boolean function $g$ obtained by substituting partial assignments $(t)$ defined by $t$ in $f$. Hence the variables in term $t$ are absent from the arguments of $g$. Thus $f/t$ is the function $g=f(S(t))$.

\subsection{Implicants and representation of satisfying assignments by partial assignments}
A sum of products (SOP) expression in $X$ is a formal Boolean sum such as
\beq\label{SOP}
F=\sum_{i}t_{i}(X)
\eeq
where $t_{i}$ are terms in $X$. By evaluation of $F$ at assignments of $X$ in $B_{0}^{n}$, $F$ corresponds to a unique Boolean function $f$. Hence we may call this an SOP representation of the function $f$ itself. A term $t$ is \emph{formally included} in an SOP form if it is present in the formal sum. An SOP form $F$ of $f$ is said to be \emph{syllogistic} if any implicant of $f$ is formally included in $F$. A \emph{consensus} of two terms $t$, $s$ in $F$ which have a single complemented literal such as $t=xu,s=x'v$ is the term $uv$ and satisfies $t+s=t+s+uv$. Given an arbitrary SOP form $F$, adding all possible \emph{consensus} terms and \emph{absorbing}\footnote{a term $t$ can be absorbed in a term $s$ if $t\leq s$} the formally included terms the resultant SOP form is syllogistic. (i.e. any implicant of $f$ is formally included in a term of this SOP). Hence this SOP form is a representation of $f$ as sum of all prime implicants $p(f)$ and it also follows that $p(f)$ is complete. This is a well known theorem \cite[theorem A.4.1]{brow} which can be stated as

\begin{theorem}\label{thbcf}\emph{A Boolean function $f(X)$ has a unique syllogistic representation
\beq\label{BCF}
f=\sum_{t\in p(f)}t
\eeq
where $p(f)$ is the set of all prime implicants of $f$.
}
\end{theorem}

This representation of $f$ in (\ref{BCF}) is known as the Blake canonical form of $f$. The uniqueness is relative to the order of prime implicant terms. However computation of prime implicants of sums and products of functions is more involved than just the SOP representation and moreover the terms in SOP representation also have the completeness property. Hence we can utilize the SOP representation to computational advantage than the Blake canonical form (\ref{BCF}). 

\subsubsection{Satisfying assignments in terms of implicants: Partial assignment representation}\label{satinimplicants}
Since our central theme is concerned with all solutions of Boolean systems we shall need a notation to represent all solutions. Let $S(f)$ denote the set of all satisfying assignments of $f$, which is the set, 
\[
S(f)=\{a\in B_{0}^{n}|f(a)=1\}
\]
which is also known as the set of all \emph{true values} of $f$. Given any SOP representation of $f$ such as (\ref{SOP}), $S(f)$ covers the union of all true values of the terms in the SOP,
\[
\bigcup S(t_{i})\subset S(f)
\]
where $S(t_{i})$ are the hypersurfaces defined by true values of the terms $t_{i}$. However when we consider the representation (\ref{SOP}) if any point $a$ in $B_{0}^n$ is in the set of true values of $f$ then $t(a)$ cannot be zero for all the terms $t$ in the SOP form (\ref{SOP}). Hence there exists a term $t$ such that $t(a)=1$. (In fact that is why there is also a prime implicant $t$ in $p(f)$ such that $t_{a}=1$). Hence the above inclusion is in fact an equality. 
\[
\bigcup S(t_{i})=S(f)
\]
This is in fact the completeness property of the set of implicants in any SOP form of $f$ and also the $p(f)$. Hence we have an alternative way to express theorem \ref{thbcf} as following proposition

\begin{prop}\label{prop:imprep}\emph{$S(f)$ is the union of hypersurfaces
\[
S(f)=\bigcup_{t\in p(f)}S(t)
\]
}
\end{prop} 
In fact the terms $t$ in the union need only be from any SOP and need not be prime implicants. Hence the search for all solutions of $f(X)=1$ need only be restricted to assignments $S(t)$ satisfying the implicants $t$ in any SOP form (\ref{SOP}) of $f$. We shall denote a complete set of implicants of $f$ defining an SOP representation (\ref{SOP}) as $I(f)$. Note that the SOP form and consequently $I(f)$ is not unique unlike the form in terms of prime implicants (\ref{BCF}). However as just observed, any $I(f)$ is sufficient to capture the set $S(f)$ as the union of true values of $t$ in $I(f)$ through partial assignments denoted as $(t)$. We can thus express $S(f)$ as

\begin{equation}\label{partialassgnform}
S(f)=\{(t_{1}), (t_{2}),\ldots,(t_{m})\}=\bigcup_{t\in I(f)}(t)   
\end{equation}
where $t_{i}$ are the implicants of $f$ in any complete set $I(f)$. $S(f)$ is empty if there are no implicants present i.e.\ when $f$ is identically zero and $S(f)=(1)=B_{0}^{n}$ when $f=1$ is a tautology. We can utilize this compact way of representing all satisfying solutions of a Boolean function to compute representation of all solutions of combinations of functions by means of appropriate algorithms as shown for products of functions below.

\subsubsection{Complete set of implicants for products of functions}
Consider now the problem of representing the satisfying set $S(h)$ of a product $h=fg$ of two functions $f,g$. If either $f\leq g'$ or $g\leq f'$ then $S(h)$ is empty. In practical problems it is required to find $S(h)$ when $h$ is known to have several factors as such a problem is equivalent to finding satisfying assignments for a system of Boolean equations. The central issue in computation of $S(h)$ is that given a decomposition of $h$ as product $h=fg$, if a complete set of implicants of one of the factors is computed how does it facilitate the computation of $S(h)$. This methodolgy thus plays an important role in the computation of $S(h)$ in terms of computation of implicants of its factors and facilitates scalable computation.

Next proposition shows representation of the satisfying set $S(h)$ of product of two functions in terms of partial assignments obtained from implicants of individual functions.

\begin{prop}\label{Partialassgnofproduct}\emph{Let $f(X)$, $g(X)$ be Boolean functions in variables $X$ and $h=fg$. Then
\[
S(h)=\bigcup_{t\in I(f)}\bigcup_{s\in I(g/t)}(t)(s)=\bigcup_{s\in I(g)}\bigcup_{t\in I(f/s)}(t)(s)
\]
}
\end{prop}

In the right hand side (RHS) of above expression if $g/t$ is a zero function for some $t$ then the prime implicant set of $g/t$ is empty. Hence the formula for $S(h)$ shows that $h$ is SAT iff $g/t$ is SAT for at least one prime implicant $t$ in $I(f)$ (alternatively iff $f/s$ is SAT for at least one prime implicant $s$ in $I(g)$).

\bpf
If $h$ is un-SAT then $h$ is identically zero and $S(h)$ is empty. If either $f$ or $g$ is identically zero the the RHS is empty hence the relation holds. If $S(h)$ is empty but none of $f$, $g$ are identically zero, consider $I(f)$ which is always nonempty for a nonzero $f$. Since
\[
S(f)=\prod_{t\in I(f)}(t)
\]
and since $h=1$ is un-SAT, $S(t)$ have no common points with $S(g)$ for all $t$ in $I(f)$. hence $g/t$ is identically zero for all $t$. Hence the RHS is also empty.

Next we prove the identity when $h=1$ is SAT. Let $a$ belong to $S(h)$. Then $a\in S(f)\cap S(g)$. Hence there exist $t\in I(f)$, $s\in I(g)$ such that $t(a)s(a)=1$ which implies one of the two expressions on RHS. Conversely if $t\in I(f)$ then $h=1$ is SAT only if and $g/t$ is SAT. Hence $s\in I(g/t)$ along with $t$ determine a solution $(t)(s)$ by partial assignments which belongs to $S(h)$. Hence the identity holds when $h=1$ is SAT. 
\epf

We consider an example to illustrate the above method of representing all solutions by partial assignments.

\begin{example}\emph{Let $F=fg$ where $f=wx'+w'y$, $g=w'x'+wx'y+x'yz'+y'z'$
By consensus, $f=wx'+w'y+xy$. Since no further consensus and absorption exists, the set of all prime implicants are
\[
p(f)=\{t_{1},t_{2},t_{3}\}=\{wx',w'y,xy\}
\]
we have
\[
\begin{array}{lcll}
g_{1}=g/t_{1} & = & y+yz'+y'z'=y+y'z'=y+z' & p(g_{1})=\{y,z'\}\\
g_{2}=g/t_{2} & = & x'+x'z'=x' & p(g_{2})=\{x'\}\\
g_{3}=g/t_{3} & = & 1 & p(g_{3})=\{1\}
\end{array}
\]
\[
S(F)=\{(wx')(y),(wx')(z')\}\bigcup\{(w'y)(x')\}\bigcup\{(xy)\}
\]
}
\end{example}

Yet another illustrative example is

\begin{example}\emph{let $F=fg$ where $f=w'+x+z'$ and $g=wxy'+wyz'+wx'z'$. We have $p(f)=\{w',x,z'\}$.
\[
\begin{array}{lcll}
g/w'& = & 0 & p(g/w')=\empty\\
g/x & = & wy'+wyz'=wy'+wyz'+wz'=wy'+wz' & p(g/x)=\{wy',wz'\}\\
g/z' & = & wxy'+wy = wxy'+wy+wx=wx+wy & p(g/z')=\{wx,wy\}
\end{array}
\]
hence we get
\[
S(F)=\{(x)(wy'),(x)(wz')\}\bigcup\{(z')(wx),(z')(wy)\}=\{(wxy'), (wxz'), (wyz')\}
\]
}
\end{example}
This shows that representation of all solutions in terms of implicants provides a compact way to represent exponentially growing set of solutions. Note that if instead of any complete set $I(f)$ of implicants the set of prime implicants $p(f)$ is used then the resulting set of implicants $I(fg)$ is not necessarily prime $p(fg)$. Based on the above proposition we develop parallel algorithms for representation of all solutions of Boolean systems in terms of implicants. In fact the algorithm proposed in the next section for computing $I(f)$ actually computes a special set $I(f)$ called orthogonal implicants which is complete and results in an orthogonal SOP or what is well known as orthogonal DNF representation of $f$.

\section{Implicant based solver}
In this section we develop the algorithm to find all solutions of the system (\ref{Boolsys}) and represent them in terms of a set of implicants denoting partial assignments or return an empty set when the system is un-SAT. Primarily the algorithm extends the process of representing the set $S(F)$ as in proposition \ref{Partialassgnofproduct} in terms of partial assignments specified by implicant terms.

Consider the Boolean system brought into standard form (\ref{Boolsys}). The individual equations $f_{i}=1$ are input to the algorithm. The ordering of the functions is assumed to be carried out before. Our algorithm for computing all satisfying assignments of this system is based on proposition \ref{Partialassgnofproduct}. Let
\[
F=\prod_{i=1}^{m}f_{i}
\]
then all solutions of (\ref{Boolsys}) are given by partial assignments defined by a complete set of implicants of $F$. Since $F$ has the recursive factorization defined by,
\[
\begin{array}{lcl}
F_{m} & = & f_{1}F_{m-1}\\
F_{m-k} & = & f_{k+1}F_{m-k-1}\\
F_{1} & = & f_{m}
\end{array}
\]
it follows from proposition \ref{Partialassgnofproduct} that if $I(f_{1})$ is a complete set of implicants of $f_{1}$ then a complete set of implicants of $F_{m}$ is 
\[
\{ts,s\in I(F_{m-1}/t), t\in I(f_{1})\}
\]
Hence recursively we can compute a complete set of implicants of $F$ by proceeding with the function 
\[
g_{2}=F_{m-1}/(t)
\]
for all implicants $t$ in $I(f_{1})$. This processes is then repeated. During this process if any of the ratios $F_{m-1}/(t)$ result in a contradiction (i.e. is zero), then the equations are not consistent and the system has no solutions with the partial assignments which satisfy the implicant. The process is then repeated for each $t$ to finally get all solution assignments. This broad algorithm (BooleanSolve below) is described here. Note that the factors $f_{i}$ of a Boolean formula $F$ obtained from the system (\ref{Boolsys}) can be chosen in the above process in any order. We call the chosen function at a step a \emph{pivot}. The function GenerateImplicant generates a complete set of implicants of a chosen pivot in the algorithm given later after this broad algorithm. 

\subsection{Main algorithm with parallel threads}
The main algorithm explained above is presented below. The algorithm has independent threads to be computed which are relegated to independent parallel processes. A thread begins at an implicant $t$ chosen from the first set $X$ which denotes set of implicants of the first pivot. At each such $t$ an independent thread thread segment begins with input $t$ and a reduced formula $F$ from the previous computation. The overall algorithm is expressed recursively due to the natural recursive structure available.

\begin{algorithm}
\DontPrintSemicolon
\KwIn{A formula $F=\prod f_{i}$ with Boolean functions $f_{i},i=1,\ldots,m$ as factors}
\KwOut{A set of implicant terms $I(F)$ which represent partial assignments of variables of all satisfying assignments of $F$. Returns empty set if $F$ is not satisfiable}
$I=\emptyset$\;
\If{$f_{i}=0$ for some $i\in [1,m]$}{Return $I(F)=I$, Go to end}
\If{$m=1$}{Return $I(F)=\mbox{GenerateImplicant}(f_{1})$, Go to end}
\Else{Choose a pivote factor $f$ from $f_{i},i\in [1,m]$\;
$X=\mbox{GenerateImplicant}(f)$\;
\For{each $t\in X$ \% start a new thread with input $(F,t)$}
{
$f_{i}\gets f_{i}/t$\;
find $m$ after deleting $f_{i}=1$\;
$F\gets \prod_{i=1}^{m}f_{i}$\;
$Y=\mbox{BooleanSolve}(F)$ \% Recursive call\;
$I\gets t\times Y$ \% the set of products $\{ts,s\in Y\}$\;
}
\textbf{end For}\;
Return $I$ \% End thread
}
\textbf{end}\;
Gather thread outputs $I$, $I(F)$ is the unioun of all thread outputs\;
\caption{BooleanSolve}
\label{ImplicantSolver}
\end{algorithm}

The function GenerateImplicant(f) is called to generate an orthogonal set of implicants of a pivot function. The pivot function may be arbitrary, however a careful use of pivot affects time of computation drastically. A general heuristic is to use the pivot function which has smallest (or sufficiently small) number of variables. 

\subsection{Implicant computation}
Computation of a complete set of implicants of Boolean functions is equivalent to computation of an SOP form. There is no unique SOP form since the minterm canonical form as well as the BCF discussed in theorem \ref{thbcf} are both SOP forms. As discussed in subsection \ref{satinimplicants} all we need to find to get all satisfying solutions of Boolean functions is to have complete set of implicants which is any set of terms appearing in an SOP form of the function. Such a computation thus avoids computation of consensus required in arriving at the prime implicants. We develop a procedure to compute an SOP form for a function in ANF.

This procedure utilizes orthonormal systems of variables. Consider an ordered set of variables $X=\{x_{1},x_{2},\ldots,x_{n}\}$ then the set of $n+1$ terms $\{t_{0},t_{1},\ldots,t_{n}\}$ given by
\[
\begin{array}{lcl}
t_{i} & = & x_{0}'x_{1}'\ldots x_{i}'x_{i+1} \mbox{ for }i=1,\ldots,n-1\\
t_{n} & = & x_{0}'x_{1}'\ldots x_{n}'
\end{array}
\]
is an orthonormal (ON) set of terms since it satisfies $t_{i}t_{j}=0$ for $i\neq j$ and
\[
\sum_{i=0}^{n}t_{i}=1
\]
We now reproduce a well known theorem \cite{brow} on orthonormal expansion.

\begin{theorem}\label{ONexpansion}\emph{If $f(X)$ is a Boolean function in variables $X$ and $\{\phi_{1},\phi_{2},\ldots,\phi_{m}\}$ is an ON set of functions then $f(X)$ has representation
\[
f(X)=\sum_{i=1}^{m}\alpha_{i}(X)\phi_{i}(X)
\]
where $\alpha_{i}(X)$ are any functions which satisfy the realtion
\beq\label{falpharel}
f(X)\phi_{i}(X)=\alpha_{i}(X)\phi_{i}(X)
\eeq
}
\end{theorem}

An important consequence of this theorem for the present computation of complete set of implicants of $f(X)$ is given by following corollary proved in \cite{desu}.

\begin{cor}\emph{There exists a satisfying assignment $a$ of $f(X)$ in $B_{0}^n$, iff there exists a unique $i$ such that $\alpha_{i}(a)=\phi_{i}(a)=1$.
}
\end{cor}

This corollary shows that a complete set of implicants can be obtained as the union of all satisfying assignments of the equations $\alpha_{i}(X)=\phi_{i}(X)=0$ for all $i$. This approach becomes useful in computation of the set of implicants if the assignments $a$ can be easily found. Such is the case when $\phi_{i}$ are orthonormal terms, since $a$ is the unique partial assignment for a term $t$ given by $(t)$. Further, $\phi_{i}(X)$ satisfies the relation with $f$ and $\alpha$ in the theorem which shows that
\beq\label{alphainONex}
\alpha_{i}(a)=(f/\phi_{i})(a)
\eeq
Hence an actual computation of the function $\alpha_{i}(X)$ is not required to compute the implicants. Note that the resulting implicants are also orthogonal. This fact follows from the expansion used in terms of orthogonal terms. As an implicant discovered is always an orthogonal term $\phi_{i}$ and further implicant of $\alpha_{i}$ is also obtained in terms of members of an ON set, it follows that the products of these implicants are also orthogonal. We can formalize this as the next important proposition (actually deserves to be a corollary of theorem \ref{ONexpansion}) which we use later to develop an algorithm for computing orthogonal implicants of a function.

\begin{prop}\emph{Let $T=\{t_{1},t_{2},\ldots,t_{r}\}$ be an ON set of terms in variables $X$ and a Boolean function $f(X)$ is expanded as in theorem \ref{ONexpansion}. Then one of the following conditions hold.
\begin{enumerate}
\item If for some $i$ the ratio $f/t_{i}=0$ then $t_{i}$ is not an implicant of $f$.
\item If $f/t_{i}=1$ then $t_{i}$ is an implicant of $f$.
\item If $f/t_{i}$ is a function of variables $X\setminus X_{i}$ where $X_{i}$ are variables in $t_{i}$, and $s$ is an implicant of $f/t_{i}$ in the new variables then $ts$ is an implicant of $f$.
\end{enumerate}
}
\label{OGproductimp}
\end{prop}

\bpf
First two assertions are obvious. For the third, note the relation (\ref{alphainONex}) which is equivalent to the fact $\alpha_{i}=f/t_{i}$. Hence if $s$ is an implicant of $\phi_{i}$ then from the relation (\ref{falpharel}) between $f$, $\alpha_{i}$ and $t_{i}$ in the ON expansion of $f$, it follows that at the partial assignment $(s)$ at an implicant $s$ of $\alpha_{i}$, denoting the set of assignments $\{s=1\}$
\[
f(s=1)\alpha_{i}(s=1)=\alpha_{i}(s=1)t_{i}(s=1)
\]
But $t_{i}$ is free of variables involved in $\phi_{i}$ hence in terms of partial assignments $(s)$ and $(t)$ we get
\[
f(t=1,s=1)=1
\]
Hence $st$ is an implicant of $f$.
\epf

Above proposition shall be used in the next section for developing a recursive algorithm for computing an OG implicant set of $f$. We first show the utility of this approach in an example.

\begin{example}\emph{Consider the function
\[
f=1\oplus w\oplus x\oplus z\oplus wy\oplus wz\oplus xz\oplus wxy\oplus xyz\oplus wyz
\]
Start with the ON set of terms $T=\{t_{i},i=0,1,2,3,4\}$
\[
T=\{w,w'x,w'x'y,w'x'y'z,w'x'y'z'\}
\]
$f/t_{4}=1$. Hence $t_{4}$ is an implicant of $f$. Next $f/t_{3}=0$ which is a contradiction. Hence $t_{3}$ is rejected. Further, 
\[
f/t_{2}=1\oplus z
\]
Hence $f/t_{2}=1$ implies $z=0$ hence we get the implicant $t_{2}z'=w'x'yz'$. Next
\[
f/t_{1}=1\oplus 1\oplus z\oplus z\oplus yz=yz
\]
Hence we have the implicant $t_{1}yz=w'xyz$. Finally 
\[
\begin{array}{lcl}
f_{1}=f/t_{0} & = & 1\oplus 1\oplus x\oplus z\oplus y\oplus z\oplus xz\oplus xy\oplus xyz\oplus yz\\
 & = & x\oplus y\oplus xz\oplus xy\oplus xyz\oplus yz
\end{array}
\]
We now start with a reduced ON set in variables $\{x,y,z\}$,
\[
S=\{s_{0},s_{1},s_{2},s_{3}\}=\{x,x'y,x'y'z,x'y'z'\}
\]
and compute $f_{1}/s_{i}$.
\[
f_{1}/s_{3}=0,\mbox{ }f_{1}/s_{2}=0,\mbox{ }f_{1}/s_{1}=1\oplus z
\]
Hence we have an implicant for $f$ as $t_{0}s_{1}z'=wx'yz'$. Finally 
\[
f_{1}/s_{0}=1\oplus y\oplus z\oplus y\oplus yz\oplus yz=1\oplus z
\]
which has the implicant $z'$ hence the implicant for $f_{1}$ is $xz'$ and that for $f$ is $wxz'$. This gives us the a complete set of implicants of $f$ as
\[
\{wxz', wx'yz', w'xyz, w'x'yz', w'x'y'z'\}
\]
Hence we also have an SOP representation of $f$ as,
\[
f=wxz'+wx'yz'+w'xyz+w'x'yz'+w'x'y'z'
\]
There are no terms with opposition. Hence consensus doesnt exist between any terms. Also there is no absorption. Hence these implicants are also prime and the SOP form is also BCF. In this case the implicants are also orthogonal since they were obtained from orthogonal impicants at the start. 
}
\end{example}

\subsubsection{Algorithm for generating orthogonal implicants, GenerateImplicant()}
From proposition \ref{OGproductimp} we can develop the following algorithm for generating a complete set of implicants of a function $f$ which are in addition also orthogonal. See algorithm \ref{implicant}. Main step of this algorithm is to first choose a term $t$ from an ON set of terms in variables $X$ of the function $f$. Then find the quotient $q=f/t$. There are now three possibilities, according to proposition \ref{OGproductimp}, 1) $q=0$. Then $t$ cannot be an implicant of $f$ hence discard $t$ and choose the next term from the orthogonal set. 2)$q=1$, Hence $t$ is an implicant. 3) $q$ has reduced set of variables left after substituting $t=1$ assignment in $X$. Hence we find an implicant $s$ of $q$ in these new variables and take the product $ts$ as the implicant. Next we update $t$ and repeat the process. This is described by following pseudocode. Now in the recursion each time we choose an ON set of terms for generating implicants. Hence if an implicant $t$ is chosen and $s_{1}$, $s_{2}$ are OG implicants in the remaining variables then $ts_{1}$, $ts_{2}$ are also OG. Hence this algorithm results into an OG set of implicants of $f$. 

\begin{algorithm}
\DontPrintSemicolon
\KwIn{$f$ a Boolean function (or an equation $f=1$)\% whose complete set of implicants is to be found}
\KwOut{$I$ a complete set of implicants of $f$ (or an equation $f=1$)}
$X=\mbox{var}\,(f)$\;
Compute $ON(X)$\% generates an orthonormal set in variables $X$, an ordered set\;
$I=\emptyset$\;
Choose first term $t$ in $ON(X)$\;
\Repeat{all terms in $ON(X)$ are processed}
{ 
Compute $q=f/t$\;
\If{$q=1$}{$I\gets I\bigcup \{t\}$}
\Else{
     \If{$q=0$}{go to end}
           \Else{
                 $I_{\rm temp}=\mbox{GenerateImplicant}(q)$\;
                 $I\gets I\bigcup \{(t)\times I_{\rm temp}$\}\;
                 \% Product set of implicants $t$ with each term of $I_{\rm temp}$.
                 }    
 \Return $I$\;
}
\textbf{end}\; 
Update $t$ to next term in $ON(X)$\;
}
\caption{GenerateImplicant}
\label{implicant}
\end{algorithm}

\subsubsection{Notation for ANF, terms and partial assignments} 
An illustrative example is now presented for demonstrating the above algorithm. A notation for representing Boolean functions in ANF is utilized which is well known in SAGE software. In this notation a Boolean function in variables $x_{i}$ is represented by an array of arrays. Consider a Boolean function
\[
f(x_{1},\ldots,x_{n})=a_{0}\bigoplus_{j}a_{j}x_{j}\bigoplus_{(i,j)}a_{ij}x_{i}x_{j}
\bigoplus_{(i,j,k)}x_{i}x_{j}x_{k}\ldots
\]
Then it is presented as an array of arrays such that the first term is $1$ when $a_{0}=1$ else absent. Then the linear terms are represented by one element subarrays $[i]$ whenever $a_{i}=1$ otherwise absent. The quadratic terms are represented as subarrays of two indices $[i,j]$ whenener $a_{i,j}=1$ or else these are absent. Similarly qubic and higher orfer terms are represented by subarrays of multiple indices whenever the co-efficients are equal to $1$ other wise they are absent. Next we denote terms such as $t=x_{1}x_{2}x_{3}'x_{4}$ by $(1,2,-3,4)$ as well as the partial assignment defined by $t$. Product of two terms $(1,2,-3)$ and $(2,-3,4)$ is the term $(1,2,-3,4)$ while that of $(1,-3,4)$ and $(2,3)$ is the empty term $()$. We write an SOP $x_{1}'x_{2}x_{3}'+x_{2}'x_{3}$ as $(-1,2,-3)+(-2,-3)$.

\begin{example}\emph{The function
\[
f(x_{1},x_{2},x_{3})=1\oplus x_{1}\oplus x_{3}\oplus x_{1}x_{2}\oplus x_{2}x_{3}\oplus x_{1}x_{2}x_{3}
\]
in above notation is represented by the array of arrays
\[
[1,[1],[3],[1,2],[2,3],[1,2,3]]
\]
Similarly
\[
f(x_{1},x_{2},x_{3},x_{4})=x_{4}\oplus x_{1}x_{3}\oplus x_{2}x_{4}\oplus x_{1}x_{3}x_{4}\oplus x_{2}x_{3}x_{4}
\]
is represented by
\[
[[4],[1,3],[2,4],[1,3,4],[2,3,4]]
\]
}
\end{example}

\subsubsection{Example illustrating the algorithm}
Next we present an example to illustrate the algorithm using above notation.

\begin{example}\emph{
Consider the Boolean formula $F$ where factors $f_{i}$ are represented in above notation. We find all satisfying assignments of $F$ if they exist or else return an empty set
\[
\begin{array}{rcl}
f_{1} & = & [[1],[2],[2,3]]\\
f_{2} & = & [[2],[3],[3,4]]\\
f_{3} & = & [[3],[4],[4,1]]\\
f_{4} & = & [[4],[1],[1,2]]
\end{array}
\]
First step, we choose $f_{1}$ as pivot and find its OG SOP. An ON set in the variables is
\[
\{(1),(-1,2),(-1,-2,3),(-1,-2,-3)\}
\]
$f_{1}/(-1,-2,-3)=0$ hence discard $(-1,-2,-3)$. $f_{1}/(-1,-2,3)=0$ hence discard $(-1,-2,3)$.
\[
\begin{array}{rcl}
f_{1}/(-1,2) & = & [1,[3]]\Rightarrow (-1,2,-3)\in I(f_{1})\\
f_{1}/(1) & = & [1,[2],[2,3]]
\end{array}
\]
Rewrite $f_{1}\leftarrow f_{1}/(1)$. Take the ON set $\{(2),(-2,3),(-2,-3)\}$. $f_{1}/(-2,-3)=1$ hence $(1,-2,-3)$ is an implicant. Also $f_{1}/(-2,3)=1$ hence $(1,-2,3)$ is also an implicant. $f_{1}/(2)=[[3]]$ hence $(1,2,3)$ is an implicant. This gives a complete OG set $I(f_{1})$ and the SOP
\[
f_{1}=(-1,2,-3)+(1,-2,-3)+(1,-2,3)+(1,2,3)
\]
Now reduce the formula $F$ taking ratios with these implicants. This causes four independent threads. Implicant chosen $(-1,2,-3)$.
\[
\begin{array}{rcl}
f_{2}\leftarrow f_{2}/(-1,2,-3) & = & [1] \\
f_{3}\leftarrow f_{3}/(-1,2,-3) & = & [[4]] \\
f_{4}\leftarrow f_{4}/(-1,2,-3) & = & [[4]]
\end{array}
\]
Hence $(-1,2,-3,4)$ is an implicant of the system. Thus thread with implicant $(-1,2,-3)$ ends in an answer in a thread with one segment length.
\[
\begin{array}{rcl}
f_{2}\leftarrow f_{2}/(1,-2,-3) & = & 0
\end{array}
\]
Hence reject this implicant. Choose next $(1,-2,3)$ and reduce $F$,
\[
\begin{array}{rcl}
f_{2}\leftarrow f_{2}/(1,-2,3) & = & [1,[4]] \\
f_{3}\leftarrow f_{3}/(1,-2,3) & = & [1] \\
f_{4}\leftarrow f_{4}/(1,-2,3) & = & [1,[4]]
\end{array}
\]
This system leads to a solution assignment $(-4)$. Hence an implicant and solution partial assignment is $(1,-2,3,-4)$. This thread also results in a one length thread with an implicant of $F$. Finally choose the implicant $(1,2,3)$.
\[
\begin{array}{rcl}
f_{2}\leftarrow f_{2}/(1,2,3) & = & [[4]] \\
f_{3}\leftarrow f_{3}/(1,2,3) & = & [1] \\
f_{4}\leftarrow f_{4}/(1,2,3) & = & [[4]]
\end{array}
\]
Thus we have an implicant $(4)$. Hence the solution partial assignment is $(1,2,3,4)$. Hence all satisfying solution partial assignments (actually full assignments) of the formula $F$ are
\[
\{(-1,2,-3,4),(1,-2,3,-4),(1,2,3,4)\}
\]
These are just three points. All threads which result into an implicant of $F$ are parallel and have length one segment.
}
\end{example}

\subsection{Time complexity and parallel speedup}
The algorithm BooleanSolve \ref{ImplicantSolver} principally uses two main computations. Algorithm GenerateImplicant \ref{implicant} for computing all implicants of a pivot and the substitution of implicants of the pivot into the remaining formula. Once implicants of a pivote are generated further computations are performed in independent threads. Thus the number of thread segments grows exponentially at the start of the algorithm. However at the end of each thread segment the substituion of an implicant results in a reduced formula from which a new pivot is chosen in and the number of new threads generated is proportional to the size of the new set $X$. The analysis of space and time complexities can thus be carried out as follows.

\subsubsection{Time taken with infinite parallel resource}
    Assuming that a sufficient parallel processing resource is available, the time taken by main algorithm equals the largest time taken by a thread which ends either in a solution or without a solution. Each such thread goes through a longest sequence of thread segments which are unit computations essentially that of reducing the formula after factoring the pivot by substituion. The new implicants that are discovered are appended to earlier implicant being processed at the start of a thread segment. Hence a largest number of thread segments are required sequentially when each thread segment discovers only one literal as implicant or rejects the implicant.
    
    During each thread segment only a single factor of $F$ is updated at the start of the thread. This factor, the pivot is processed for generating implicants and a single implicant is substituted in all other factors. Hence if the factors $f_{i}$ in the formula $F$ are sparse (or the system of equations is sparse), which implies each factor has small number of variables as compared to $n$ and further we assume that a bound on the number of variables present in each $f_{i}$ is fixed, then the time taken to generate implicants of $f_{i}$ is small and an upper bound can be taken as a constant. This time in fact reduces in every sequential step as assignments of variables are found. Hence substitution of an implicant in all factors and updating the equations to delete those which result in $f_{k}/t=1$, occupies the main computation time in each thread segment. Each of these operations can be then assumed to be accomplished in a constant time. Hence for the longest thread this time is proportional to $n$ in the worst case. Hence it follows that since all thread segments are computed in parallel
    \[
    \mbox{Time required for solving $m$ equations in $n$ variables }T(n,m)=O(n)
    \]
\subsubsection{Parallel speedup}
In a sequential execution of the algorithm all threads are executed one after the other. Hence when sufficient parallel resource is available the best (or maximum) speedup achievable is
\[
S=\frac{\mbox{sum of time taken for each thread}}
{\mbox{time taken for the longest thread}}
\]
Hence speedup achieved will be greater, if number of threads generated is greater. In terms of number of thread segments we can estimate this speedup under infinite resource as follows. Let $N$ denotes the total number of thread segments. We assume that each thread segment under above conditions takes approximately a constant computation time. If the number of sequential thread segments in the longest thread is $N_{\rm seq}$ then the speedup is
\[
S=N/N_{\rm seq}
\]
Hence it follows that maximum speedup can be achieved if 
\[
p\geq (N/N_{\rm seq})
\]
However, less number of processors are actually required for handling independent threads by reusing the processors which finish their computation of a thread segment. Full details of such an algorithm shall be explored in another article.

\subsubsection{Seedup estimate based on Amdhal's law}
In the above algorithm total number of thread segments is $N$ (which we called unit operations above) and in a sequential processing these are executed in a sequence for completing a thread which either results into a solution or ends in empty solution. Of these threads there is one which takes longest number of sequential stages of thread segments which is $N_{\rm seq}$. Hence this sequential computation of thread segments is essential and cannot be parallelized. This gives the fraction $f_{\rm par}$ of the total algorithm (or the code of the algorithm) that can be parallelized as follows,
\[
f_{\rm par}=\frac{N-N_{\rm seq}}{N}
\]
Hence according to Amdahl's law the speedup $S(P)$ achievable when we use $P$ number of processors to execute the unit operations is
\[
\begin{array}{lcl}
S(P) & = & [f_{\rm par}/P+(1-f_{\rm par})]^{-1}\\
 & = & \frac{P(N/N_{\rm seq})}{(N/N_{\rm seq})+P-1}
\end{array}
\]
In particular when number of processors $P=N/N_{\rm seq}$ we have the critical speedup
\[
S_{\rm crit}=\frac{P^{2}}{P^{2}-(P-1)^{2}}
\]
Similarly $S(P)\rightarrow (N/N_{\rm seq})=S$ when $P\rightarrow\infty$ as expected. 
\section{Concluding remarks}
The problem of computing all solutions of a Boolean system of equations is considered as that of finding a complete set of implicants of a Boolean formula $F=\prod f_{i}$ where $f_{i}$ are Boolean functions. The algorithm proposed has complexity bound of $O(n)$ if sufficient number of parallel processors are available where $n$ is the number of variables in $F$. (This estimate is justified only when the functions $f_{i}$ are sparse and have an upper bound on number of variables involved). The algorithm generates parallel threads starting from implicants of the a chosen (pivot) factor and reducing the system by each implicant by computing the quotients of remaining factors by implicants. The algorithm is naturally recursive. Initial rise in number of threads follows an exponential growth but soon the threads gets discarded as the quotients result into contradictions or result into a partial solution assignment. All such implicants fixing partial assignments representing solutions satisfying $F$ are thus generated. A clear understanding of the bound on total number thread segments does not seem easy to determine and shall be the crux of complexity analysis of the algorithm. Many heuristics can be followed in choice of the pivot factor for generating the implicants and decomposition of the formula $F$ into simpler forms which can be independently solved. The intermediate step of computing implicants is also a recursive algorithm and has inherent parallel structure. It is expected that the algorithm presented in this paper shall be found useful for a variety of applications of Boolean satisfiability for practically relevant cases.

\begin{center}
Acknowledgements\\
Supported by the project grant 11SG010 of IRCC of IIT Bombay. Author is thankful to Bharat Adsul for valuable help and Anmol Yadav for pointing out errors in previous versions.
\end{center}

\end{document}